\documentclass[preprint,aps,amsmath,amsfonts]{revtex4}

\usepackage{graphicx}

\begin{document}

\setlength\arraycolsep{2pt}  
\def\Pcm#1{{\mathcal{#1}}}
\def\nn{\nonumber}
\def\er#1{eqn.\eqref{#1}}
\def\fgref#1{fig.\ref{#1}}
\newcommand{\re}{\Re \textup{e} \ }
\newcommand{\im}{\Im \textup{m} \ }
\newcommand{\td}{\textup{d}}
 
\title{Butterfly Tachyons in Vacuum String Field Theory}
\author{Peter~Matlock}
\email{pwm@sfu.ca}
\affiliation{Department of Physics, Simon Fraser University, Burnaby BC, Canada}

\begin{abstract}
  We use geometrical conformal field theory methods to investigate 
tachyon fluctuations about the butterfly projector state in Vacuum 
String Field Theory. We find that the on-shell condition for the tachyon
field is equivalent to the requirement that the quadratic term
in the string-field action vanish on shell. This further motivates
the interpretation of the butterfly state as a D-brane. We begin 
a calculation of the tension of the butterfly, and conjecture that
this will match the case of the sliver and further strengthen this 
interpretation.
\end{abstract}
\maketitle 
\section{Introduction}
There are now several known solutions \cite{KP,RSZ2,GRSZ,RSZ6} to the equations of motion of
Vacuum String Field Theory (VSFT) \cite{RSZ5}. These include the sliver state \cite{KP,RSZ2,RZ} and the 
butterfly state \cite{GRSZ}. The sliver state was conjectured to represent a 
D$25$-brane, and subsequent calculations of its tension based on this
assumption yielded the correct brane tension. The sliver state construction
was used also to build solutions corresponding to D$p$-branes of arbitrary
dimension \cite{RSZ4}, and ratios of tensions were found \cite{Okuyama,MRVY} which were in agreement with
the known results from string theory. These calculations were based on a
field expansion of fluctuations about the classical solution, using
a tachyon field to find the brane tension. One requirement for the consistency of the
interpretation of the sliver brane as a D-brane is that the equation of motion
for the tachyon must be a consequence of the string field equation of
motion; an on-shell string field must correspond to an on-shell tachyon.
This means that the quadratic term in the resulting tachyon action must vanish,
as is the case for the sliver state.\cite{Okawa}
While it has already been assumed in the literature that the butterfly
can be interpreted as a D-brane, this may have been premature, as the 
above properties had not been verified. Were the tachyon field not to satisfy these
requirements, it would mean that the butterfly state, although known to be a brane 
(\emph{i.e.}~localised)
solution and a rank-one projector, could not be viewed as a D-brane.
We show in the present paper that the butterfly does indeed support
a tachyon field with vanishing on-shell quadratic term.

We now turn our attention to the brane tension.
The tension of the brane may be obtained from the cubic term in the tachyon 
action \cite{Okuyama,Okawa,MRVY}, and this has been completed for the case of the 
sliver by Okawa \cite{Okawa}. In that calculation, the evaluation of the cubic terms 
turned out to be very tedious and lengthy, and in the present case of the butterfly 
we find that it is much more so. Thus rather than attempting a calculation of the entire
cubic term, we content ourselves with motivating the procedure and conjecture
how the result should obtain. In this way, we will see how the correct brane 
tension should arise.

This paper is structured as follows. In section \ref{BS} we review the
geometrical construction of the butterfly state as a surface state \cite{RSZ5},
along with the regularisation required for any concrete calculations.\cite{GRSZ}
We then in section \ref{TF} turn to the expansion of deformations of a VSFT solution \cite{Okawa}.
In section \ref{quad} we investigate the quadratic term in the tachyon action.
This surface-state calculation involves the construction of conformal mappings
in order to perform star multiplication.
In section \ref{tens} we begin the calculation of the ratios of tensions of
different butterflies, suggest how this could be completed, and comment on 
the preliminary results.
We conclude in section \ref{DIS} with a discussion of surface states and
the r\^ole of regularisation and conformal invariance with respect to deformations
of string fields and definition of fields.

\section{The Butterfly State}
\label{BS}
We use the geometrical, surface representation of string fields; thus 
we specify states using a BPZ product
with an arbitrary state $|\phi\rangle$. For details of this representation
and methods thereof, we refer the reader to \cite{RSZ5}.

\begin{figure}[ht]
\includegraphics{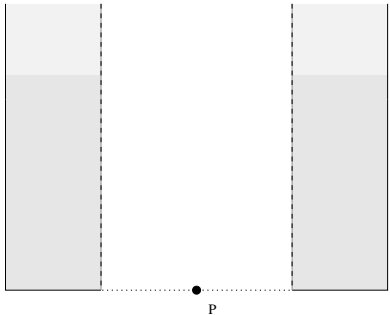}
\caption{\label{bfly1} the butterfly, defined on the surface $\Sigma$}
\end{figure}

The butterfly state $|\Pcm{B}\rangle$ is a factorisable state, so that it 
may be decomposed into 
the product of a left-string functional and a right-string functional.
The surface $\Sigma$, defined by $-\pi/2 < \re z < \pi/2$ and $\im z>0$, used to 
define the butterfly is shown in \fgref{bfly1}.
 The unshaded 
region is the local patch, $-\pi/4 < \re z < \pi/4$, $\im z>0$. The dashed lines which 
border the local patch are the left- and right-string boundaries, and the solid line is 
the boundary of the 
surface on which we impose the standard open string boundary condition.
In the centre of the local patch is the puncture $P$ where we insert the operator $\phi$,
transformed to this coordinate system from the canonical half-disc via the mapping $f$.
Thus the BPZ product of the butterfly with the arbitrary state represented by the 
operator $\phi$ is
\begin{equation}
 \langle \Pcm{B} | \phi \rangle = \langle f \circ \phi(P) \rangle_{\Sigma}
,\end{equation}
where the combination of the mapping $f$ and the surface $\Sigma$ really 
only need be defined up to conformal equivalence. This is not strictly true of
the states we will define in subsequent sections, due to regularisation of operator
short-distance singulatities.

We will also have need of the regularised butterfly $|\Pcm{B}_h\rangle$, and we borrow the 
formulation from \cite{GRSZ}.
\begin{figure}[ht]
\includegraphics{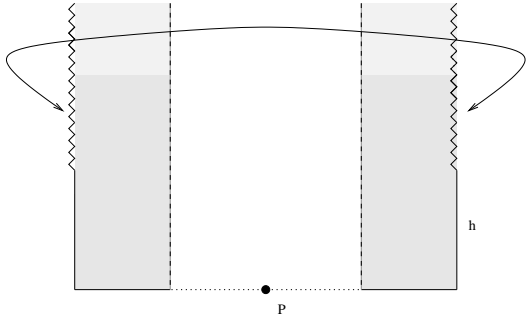}
\caption{\label{bfly2} the regularised butterfly, defined on $\Sigma_h$}
\end{figure}
The surface $\Sigma_h$, shown in \fgref{bfly2}, is obtained from $\Sigma$ by identifying the left and right edges, above
some height $h$. This is the regularisation parameter, and the limit $h\rightarrow\infty$ will
be taken to obtain the surface $\Sigma$. The state is now defined by
\begin{equation}
 \langle \Pcm{B}_h | \phi \rangle = \langle f_h \circ \phi(P) \rangle_{\Sigma_h}
.\end{equation}
 The regularised state $|\Pcm{B}_h\rangle$ does not satisfy the 
string-field equation of motion for finite $h$.
 
\section{Tachyon Fluctuations}
\label{TF}
In \cite{Okawa} an elegant proposition was made regarding both the 
parametrisation of string-field fluctuations by fields and the
construction of states representing the coefficients of these fields.
We here present this briefly, referring the reader to that paper for 
details.

The string-field action is given by
\begin{equation}
\label{SFact}
S=-\frac12\langle\Psi|\Pcm{Q}|\Psi\rangle -\frac13\langle\Psi|\Psi * \Psi \rangle
.\end{equation}
Since the VSFT operator $\Pcm{Q}$ is purely ghost, there are factorisable 
solutions $\Psi=\Psi_m\otimes\Psi_g$ satisfying
\begin{eqnarray}
\Pcm{Q}\Psi_g + \Psi_g * \Psi_g &=& 0 \\
\Psi_m * \Psi_m &=& \Psi_m
.\end{eqnarray}
As the ghost part of the solution is thought to be in some sense universal \cite{Sen2},
attention has mainly been given to the matter part of the solution.
From \cite{Okawa}, a finite deformation of the matter solution
parametrised by fields $\{\varphi_i\}$ is given by
\begin{equation}
\label{defo1}
\langle\Psi_{\{\varphi_i\}}|\phi\rangle=
\Pcm{N}\left\langle
              \exp \left[
                   -\int_{\tilde{\partial}\Sigma} \td z  \int \td k \sum_i 
                  \varphi_i(k)\Pcm{O}_{\varphi_i(k)}(z)
                \right]f\circ\phi(P)\right\rangle_\Sigma
,\end{equation}
where $\tilde{\partial}\Sigma$ refers to the portion of the boundary of $\Sigma$ belonging to
the state $|\Psi_{\{\varphi_i\}}\rangle$, as opposed to the reference state $|\phi\rangle$.
${\{\varphi_i\}}$ are fields which parametrise the deformation, while $\Pcm{O}_{\varphi_i(k)}$
are the corresponding vertex operators. The integral of such a vertex operator, which 
is of conformal dimension one for on-shell physical states, is thus conformally invariant.

In the case of a tachyon deformation, we have
\begin{equation}
\langle e^{-T}|\phi\rangle=
\Pcm{N}\left\langle
              \exp \left[
                   -\int_{\tilde{\partial}\Sigma} \td z  \int \td k T(k)e^{-ik\cdot X}(z)
                \right]f\circ\phi(P)\right\rangle_\Sigma
.\end{equation}
Expanding in powers of the tachyon field $T$, we have
\begin{equation}
\label{FlExp}
\langle e^{-T}|\phi\rangle=\sum_j \langle T_j |\phi\rangle
,\end{equation}
where
\begin{equation}
\label{Tj}
\langle T_j |\phi\rangle = \left\langle \frac1{j!}\left( -\int \td z \int \td k T(k) e^{-ik\cdot X}(z) \right)^j f\circ\phi(P) \right\rangle_\Sigma
.\end{equation}
One must take care to regularise these states when taking BPZ products, as short-distance 
singularities will obtain. 

$|T_0\rangle$ is nothing but the classical solution $|\Psi_m\rangle$. The next term in the series \eqref{FlExp} is
\begin{equation}
\label{TachExp}
|T_1\rangle = -\int \td k T(k)|\chi_T(k)\rangle
,\end{equation}
where the tachyon state $|\chi_T(k)\rangle$ is given by the intergral of the 
tachyon vertex operator along the boundary of the surface.
The linearised equation of motion for the tachyon state is then
\begin{equation}
\label{tyeom}
|\chi_T(k)\rangle = |\chi_T(k) * \Psi_m \rangle + |\Psi_m * \chi_T(k)\rangle
.\end{equation}
It is shown in \cite{Okawa} that in the case of the sliver state $\Psi_m=\Xi_m$
this equation is satisfied on-shell not only for the tachyon $\chi_T(k)$
but for all physical string states $|\chi_{\varphi_i}(k)\rangle$.
In the case of the butterfly, it is easy to see that the linearised equation 
of motion \er{tyeom} is satisfied.

Inserting the expansion \eqref{FlExp} into the action \eqref{SFact}, the term quadratic 
in the tachyon field is immediately given by
\begin{eqnarray}
\label{quadt}
S^{(2)}&=&-\langle\Psi_g|\Pcm{Q}\Psi_g\rangle\left(
\frac12\langle T_1|T_1\rangle
+\langle T_2|T_0\rangle
-\langle T_1|T_1*T_0\rangle
-\langle T_2|T_0*T_0\rangle
\right) \\
&=&-\frac{\Pcm{K}}2(2\pi)^{26}\int\td k K(k^2)T(k)T(-k)
,\end{eqnarray}
where $K(k^2)$ consists of four pieces coming from the above four contributions;
\begin{equation}
\label{Kexp}
\frac12K(k^2)=\frac12K_{11}+K_{20}-K_{110}-K_{200}
.\end{equation}
In \cite{Okawa}, $K(k^2)$ was found to be identically zero in the case of the sliver state.
In the following section we perform an analogous calculation for the case of the butterfly.

Since we wish to deal with tension in section \ref{tens}, we will also need the 
part of the action cubic in the tachyon, given by
\begin{eqnarray}
\label{cubt}
S^{(3)}&=&-\langle\Psi_g|\Pcm{Q}\Psi_g\rangle
\bigg(
\langle T_3|T_0\rangle
-\langle T_3|T_0*T_0\rangle
+\langle T_2|T_1\rangle
-\langle T_2|T_1*T_0\rangle
-\langle T_2|T_0*T_1\rangle
\nonumber\\
&&{}-\frac13\langle T_1|T_1*T_1\rangle
\bigg) \\
&=&-\frac{\Pcm{K}}3(2\pi)^{26}\int\td k_1 \td k_2 \td k_3
\delta(k_1+k_2+k_3) V(k_1,k_2,k_3)T(k_1)T(k_2)T(k_3)
,\end{eqnarray}
where $V$ is a function containing contributions from each of the six terms in
the action.
We consider the on-shell case, and write this quantity as
\begin{equation}
\label{Vexp}
-\frac13V=V_{30}+V_{21}-V_{300}-V_{210}-V_{201}-\frac13V_{111}
.\end{equation}
For the case of the sliver, Okawa \cite{Okawa} showed that the
first five terms together cancel; the cubic action is given solely
by the sixth term $V_{111}$ and thus so is the brane tension.
We will discuss the case of the butterfly in section \ref{tens}.

\section{Quadratic Tachyon Action}
\label{quad}
Here we calculate the quadratic term in the tachyon action, and show that it 
vanishes on-shell. From \er{quadt} we have four pieces.

\begin{figure}[ht]
\includegraphics{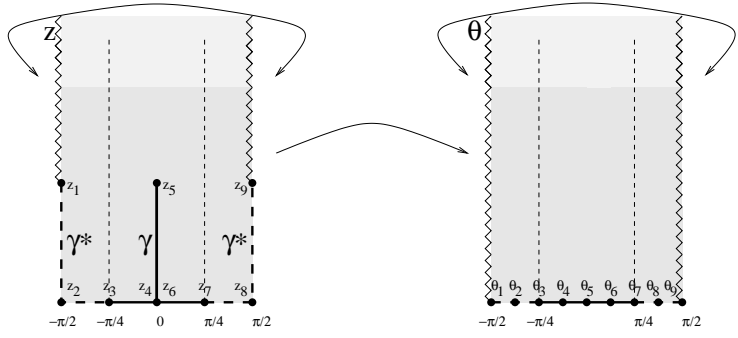}
\caption{\label{K11} mapping for two-state BPZ product} 
\end{figure}

For the term $K_{11}$, we use the mapping shown in \fgref{K11} to map
the surface obtained by gluing together two copies of $\Sigma_h$ onto a
cone of circumference $\pi$. The boundary of the first and second copies 
of $\Sigma_h$ will be denoted $\gamma$ and $\gamma *$. They are shown in 
the figure as heavy dashed and solid lines.
From the appendix, the mapping is given by
\begin{equation}
\label{maprel}
\cos 2\theta = \frac1\eta \cos 2z
,\end{equation}
where $\eta=\cosh 2h$, so that
\begin{equation}
\label{mapdif}
\frac{\td z}{\td\theta}\equiv s(\theta)=
 \frac{\varepsilon(\theta) \sin 2\theta }{\sqrt{\frac1{\eta^2}-\cos^22\theta}}
,\end{equation}
where $\varepsilon(\theta)$ denotes the sign of $\re \theta$.
From the previous section, we have
\begin{equation}
K_{11}\sim\left\langle
\int_\gamma\td z \int_{\gamma *}\td z' e^{ik\cdot X}(z)e^{ik'\cdot X}(z')
\right\rangle
,\end{equation}
where by $\sim$ we indicate that there this requires regularisation; we see
there will be short-distance singularities at $z=\pm\pi/4$.
We thus regularise the state by leaving a gap of $2\delta_z$ at each 
of these points. The limit $\delta_z\rightarrow0$ will be taken at the end
of the calculation.
Transforming to the $\theta$ coordinate, we have
\begin{equation}
K_{11}=
\int_{\theta_3+\delta_\theta}^{\theta_7-\delta_\theta} \td \theta 
\left\{
\int_{\theta_7+\delta_\theta}^{\theta_9} \td \theta' + 
\int_{\theta_1}^{\theta_3-\delta_\theta} \td \theta' 
\right\}
\left\langle
e^{ik\cdot X}(\theta)e^{ik'\cdot X}(\theta')
\right\rangle_{C_\pi}
,\end{equation}
where $C_\pi$ is a cone with opening angle $\pi$, and $\delta_\theta$ is
our regulator $\delta_z$, transformed to the theta system;
\begin{equation}
\delta_\theta=\frac1{s(\theta)}\bigg|_{\theta=\pm\pi/4}\delta_z=\frac1\eta\delta_z
.\end{equation}
The propagator on the boundary of the cone is given by
\begin{equation}
\label{coneprop}
\left\langle e^{ik\cdot X}(\theta)e^{ik'\cdot X}(\theta')\right\rangle_{C_{n\pi}}
=
\delta(k+k')\left|n\sin\frac{\theta-\theta'}{n}\right|^{2kk'}
,\end{equation}
and considering the on-shell case, $k^2=1$, we have
\begin{equation}
K_{11}=
\int_{\theta_3+\delta_\theta}^{\theta_7-\delta_\theta} \td \theta 
\left\{
\int_{\theta_7+\delta_\theta}^{\theta_9} \td \theta' + 
\int_{\theta_1}^{\theta_3-\delta_\theta} \td \theta' 
\right\}
\csc^2(\theta-\theta')
.\end{equation}
Performing the integrals, we obtain
\begin{eqnarray}
K_{11}&=&-2\log\sin 2\delta_\theta\\
      &\rightarrow&-2\log 2 -2 \log \delta_z + 2 \log \eta 
,\end{eqnarray}
where we have made use of the limit $\delta_z\rightarrow0$.
We see that there is a finite and a divergent part to $K_{11}$ for 
finite $\eta$, and also that there is a divergence 
as $\eta\rightarrow\infty$. We expect the part divergent in $\delta_z$
to cancel between $K_{11}$ and $K_{110}$ in \er{Kexp}, since products
of $|T_1\rangle$ states must be regularised at the endpoints of vertex 
operator integration, while the $|T_2\rangle$ state contains a
different, independently regularised divergence.

\begin{figure}[ht]
\includegraphics{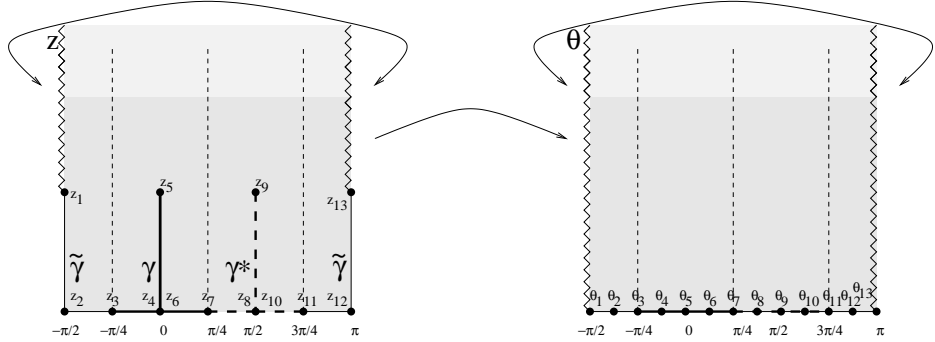}
\caption{\label{K110} mapping for three-state BPZ product} 
\end{figure}

Turning our attention now to $K_{110}$, we use the mapping shown in \fgref{K110}.
We may again write
\begin{equation}
\label{k110a}
K_{110}\sim\left\langle
\int_\gamma\td z \int_{\gamma *}\td z' e^{ik\cdot X}(z)e^{ik'\cdot X}(z')
\right\rangle
.\end{equation}
Recalling the mapping in the appendix, the relation \eqref{maprel} and
the derivative \eqref{mapdif} remain unchanged. The mapping is now to a
cone of angle $3\pi/2$, so substituting $n=3/2$ into the two-point function \eqref{coneprop},
we may write
\begin{equation}
K_{110}=
\frac49\int_{\theta_3+\delta_\theta}^{\theta_7-\delta_\theta} \td \theta 
\int_{\theta_7+\delta_\theta}^{\theta_{11}-\delta_\theta} \td \theta'
\csc^2\frac23(\theta-\theta')
.\end{equation}
Again, performing the integrals, we have
\begin{equation}
K_{110}=\frac32\log3 -3\log 2 + -\log \delta_z + \log \eta
,\end{equation}
so that the combination
\begin{equation}
\label{k1k1}
 \frac12K_{11}-K_{110}=2\log 2 - \frac32\log 3
\end{equation}
is finite, but non-zero.

In order to calculate the $K_{20}$ and $K_{200}$ contributions, greater care
must be taken with the regularisation procedure.
The $|T_2\rangle$ state from \er{Tj} involves the a double integral since
\begin{equation}
\label{di}
\left\langle\left( -\int \td z e^{-ik\cdot X}(z) \right)^2\right\rangle
=\int\td z \td z' \left\langle e^{-ik\cdot X}(z)e^{-ik\cdot X}(z')\right\rangle
,\end{equation}
but this will be divergent when $z \sim z'$. 

Looking at both \fgref{K11} and \fgref{K110}, let us calculate $K_{20}$ and $K_{200}$
simultaneously. 
We follow Okawa \cite{Okawa} and regularise the double integral \eqref{di} as
\begin{equation}
\int_{\gamma;z_3+\epsilon_z}^{z_7} \td z \int_{\gamma;z_3}^{z-\epsilon_z} 
\td z' \left\langle e^{-ik\cdot X}(z)e^{-ik\cdot X}(z')\right\rangle
,\end{equation}
where the $z$ integral is to be taken along the coutour $\gamma$
from `just after' the beginning until the end at $z_7$, and the
$z'$ integral is taken along the $\gamma$ from 
the beginning at $z_3$ until `just before' the
point $z$. That is, the small quantity $\epsilon_z$ is equal to $\epsilon$, $i\epsilon$,
$-i\epsilon$ and $\epsilon$ for each segment of $\gamma$ respectively, where real 
$\epsilon$ is then the regularisation parameter.
Using this formulation in the $\theta$-system, we write
\begin{equation}
\label{K20o}
K_{20(0)}=\int_{\theta_3+\epsilon_\theta}^{\theta_7}\td \theta
          \int_{\theta_3}^{\theta-\epsilon_\theta}\td \theta'
           \omega^2\csc^2\omega(\theta-\theta')
,\end{equation}
where
\begin{equation}
 \epsilon_\theta=\frac1{s(\theta)}\epsilon_z
\end{equation}
so that in the inner integral, the upper limit of integration depends on $\theta$.
In order to do both calculations at once, we have introduced the constant $\omega$;
for the cases of $K_{20}$ and $K_{200}$, $\omega=1$ and $\omega=2/3$, respectively.
Let us first perform the inner integral, giving
\begin{eqnarray}
\label{K2a}
K_{20(0)}&=&\int_{\theta_3+\epsilon_\theta}^{\theta_7}\td \theta
	  \left(
 		\frac1{\epsilon_\theta}-\omega\cot\omega(\theta-\theta_3)
 	\right) \\
        &\rightarrow&
       \left[\frac1{\epsilon_z} \int \td \theta s(\theta) - \log \sin \omega\theta
 	\right]_{\theta_3+\epsilon_\theta}^{\theta_7}
.\end{eqnarray}
Substituting $s(\theta)$ from \er{mapdif}, we have
\begin{equation}
\int\td\theta s(\theta)=-\frac{\varepsilon(\theta)}2
	\tan^{-1}\frac{\sin 2\theta }{\sqrt{\frac1{\eta^2}-\cos^22\theta}}
.\end{equation}
We may now evaluate \er{K2a}, obtaining
\begin{equation}
K_{20(0)}=-\frac1\epsilon
\tan^{-1}\frac{\sin 2\theta_4 }{\sqrt{\frac1{\eta^2}-\cos^22\theta_4}}
-\log\sin\frac{\omega\pi}{2}+\log\frac{\omega\epsilon}{2}
.\end{equation}
We thus find the contribution
\begin{equation}
\label{k2k2}
  K_{20}-K_{200}=-2\log2 +\frac32\log 3
\end{equation}
to \eqref{Kexp} to be again finite and non-zero, precisely cancelling the 
contribution \eqref{k1k1} to the quadratic term \eqref{Kexp}.

\section{Cubic Tachyon Action and Brane Tension} 
\label{tens}
We have verified in section \ref{quad} that the butterfly does indeed support a tachyon
field. In order to further motivate its interpretation as a D-brane state, the brane 
tension should also be calculated. This was properly done for the sliver in \cite{Okawa},
and we consider the same procedure here for the butterfly. 

We recall that the tachyon coupling constant can be identified by first canonically normalising
the tachyon field using the quadratic term, and then extracting the coefficient of the cubic term.
This coupling will be related to the tension, and the tension may be expressed in terms of the 
energy density.

The cubic term in the action contains the quantity from \er{Vexp},
\begin{equation}
\label{Vexpp}
-\frac13V=V_{30}+V_{21}-V_{300}-V_{210}-V_{201}-\frac13V_{111}
.\end{equation}
These may be calculated as follows. The three-point function on the boundary of the
cone $C_{n\pi}$ is given by
\begin{eqnarray}
P_n(w_1,w_2,w_3)\delta(k_1&+&k_2+k_3) \equiv \langle e^{ik_1\cdot X}(w_1) e^{ik_2\cdot X}(w_2) e^{ik_3\cdot X}(w_3) \rangle_{C_{n\pi}} \\
  &=&\frac1{n^3} \left| 
     \csc \frac{|w_1-w_2|}n  \csc \frac{|w_2-w_3|}n \csc \frac{|w_3-w_1|}n 
	\right|\delta(k_1+k_2+k_3)
.\end{eqnarray}
In the following, as before, we imply the use of \fgref{K11} and \fgref{K110} for two- and three-state products,
respectively. We use the same regularisation parameters $\delta_\theta=\frac1\eta\delta_z$ and 
$\epsilon_\phi\equiv\frac{1}{s(\phi)}\epsilon_z$ as in the preceding section.

We begin with $V_{111}$ and write
\begin{equation}
V_{111} \sim \left\langle
	\int_\gamma \td w_1 \int_{\gamma *} \td w_2 \int_{\tilde{\gamma}} \td w_3
        e^{ik\cdot X}(w_1) e^{ik\cdot X}(w_2) e^{ik\cdot X}(w_3) 
	\right\rangle
.\end{equation}
Regularisation may be carried out as before, so that
\begin{eqnarray}
V_{111} &=&
	\int_{\theta_3+\delta_\phi}^{\theta_7-\delta_\phi}\td \phi_1
	\int_{\theta_7+\delta_\phi}^{\theta_{11}-\delta_\phi}\td \phi_2
        \left\{
	\int_{\theta_{11}+\delta_\phi}^{\theta_{13}}\td \phi_3
	+ \int_{\theta_1}^{\theta_3-\delta_\phi}\td \phi_3
        \right\} \nonumber\\
      &&  \left(\frac23\right)^3
        \csc \frac23(\phi_1-\phi_2)
        \csc \frac23(\phi_2-\phi_3)
        \csc \frac23(\phi_3-\phi_1)
.\end{eqnarray}
The three integrals in this case are independent, (\emph{i.e.} the integration bounds do not depend on 
the integration variables) and we find that this is identical to $V_{111}$ in the case of the sliver,
calculated in \cite{Okawa}, that is
\begin{equation}
V_{111}=\frac{\pi^2}3
.\end{equation}

Moving on to $V_{21}$ and $V_{210}$ we may write the unregularised quantity as
\begin{equation}
V_{21(0)} \sim
	\int_\gamma \td w_1 \td w_2 \int_{\gamma *} \td w_3 P_{1/\omega}(w_1,w_2,w_3)
,\end{equation}
where we have written both states simultaneously using $\omega$, as we did in section \ref{quad}.
We regularise as we did with $K_{20(0)}$;
\begin{equation}
V_{21(0)} =
	2 \int_{\gamma;z_3+\epsilon_{z_3}}^{z_7} \td w_1 
          \int_{\gamma;z_3}^{w_1-\epsilon_{w_1}}\td w_2 
           \int_{\gamma *} \td w_3 P_{1/\omega}(w_1,w_2,w_3)
.\end{equation}
In the $\theta$-system, this becomes
\begin{equation}
\label{V21}
V_{21} = 2 \int_{\theta_3+\epsilon_{\theta_3}}^{\theta_7} \td \phi_1 
          \int_{\theta_3}^{\phi_1-\epsilon_{\phi_1}}\td \phi_2
	\left\{
          \int_{\theta_7+\delta_\theta}^{\theta_9} \td \phi_3
          \int_{\theta_1}^{\theta_3-\delta_\theta}\td \phi_3
	\right\}
        \csc (\phi_1-\phi_2)
        \csc (\phi_2-\phi_3)
        \csc (\phi_3-\phi_1)
.\end{equation}
Here we are using the regulator $\epsilon$ just as we did in the previous section.

$V_{210}=V_{201}$ may similarly be expressed as
\begin{equation}
\label{V210}
V_{210} = 2 \int_{\theta_3+\epsilon_{\theta_3}}^{\theta_7} \td \phi_1 
          \int_{\theta_3}^{\phi_1-\epsilon_{\phi_1}}\td \phi_2
          \int_{\theta_7+\delta_\theta}^{\theta_{11}-\delta_\theta}\td \phi_3
	\left(\frac23\right)^3
        \csc \frac23(\phi_1-\phi_2)
        \csc \frac23(\phi_2-\phi_3)
        \csc \frac23(\phi_3-\phi_1)
.\end{equation}

Finally, $V_{30}$ and $V_{300}$ may be written
\begin{equation}
V_{30(0)} \sim \int_\gamma \td w_1 \td w_2 \td w_3 P_{1/\omega}(w_1,w_2,w_3)
,\end{equation}
which may be regularised and written as
\begin{equation}
\label{V30o}
V_{30(0)} = 4 \int_{\theta_3+2\tau_{\theta_3}}^{\theta_7} \td \phi_1
	\int_{\theta_3+\tau_{\theta_3}}^{\phi_1-\tau_{\phi_1}} \td \phi_2
	\int_{\theta_3}^{\phi_2-\tau_{\phi_2}} \td \phi_3
	\omega^3
        \csc \omega(\phi_1-\phi_2)
        \csc \omega(\phi_2-\phi_3)
        \csc \omega(\phi_3-\phi_1)
.\end{equation}
Here, $\tau$ is a new regulator, which is used exactly as is $\epsilon$ in previous
expressions. This means that $\tau_\phi=\frac1{s(\phi)}\tau_z$, and $\tau_z$ `follows the contour'
in the $z$-system, as explained for $\epsilon_z$ just before equation \eqref{K20o}.

Explicit evaluation of the integrals for $V_{21(0)}$ and $V_{30(0)}$ in 
equations \eqref{V21}, \eqref{V210} and \eqref{V30o} is not impossible, but very tedious.
Primarily this is due to the complicated dependence of the upper limits of the 
inner integrals on the variables of the outer integrals. 
We leave this evaluation for future investigation and here make some comments.
We first mention that the calculation must be done for fixed, finite $\eta$, taking
the limit $\eta\rightarrow\infty$ only at the very end. The $\delta\rightarrow0$, $\epsilon\rightarrow0$
and $\tau\rightarrow0$ limits may be considered while performing the calculation,
but care must be taken to keep all divergent terms in these regulators. In addition,
in individual expressions, these three limits do not always commute, and it is only at the end
where the divergent parts should be seen to cancel. Given that these two regulators are independent,
we again expect that the combinations $V_{30}-V_{300}$ and $V_{21}-V_{210}-V_{201}$ will not be divergent.
We also expect that combined, they contribute zero to the expression \eqref{cubt} for the 
cubic tachyon coupling, through \er{Vexp}, leaving the tension dependent only on the term $V_{111}$.
Finally, to calculate the tension one must calculate the overall normalisation of the quadratic term,
so that it may be written canonically and the cubic term normalised appropriately. This calculation also
is tedious for the case of the butterfly; we do not undertake it here.
Work is in progress on the quadratic normalisation and the cubic term evaluation; if reasonable, we 
intend to address the full calculation of the tension in a future publication.

\section{Discussion}
\label{DIS}
   We have calculated the on-shell quadratic term in the tachyon action, and found that it vanishes.
The structure of this calculation is the same as that for the sliver \cite{Okawa}.
This is compatible with the assertion that the butterfly represents a D-brane state. The tension
could be calculated by following the procedure outlined in the last section, and we expect that this
will also agree with the canonical value of unity for a D-brane, matching the case of the sliver.

Finally, it is interesting to note that when defining surface states with operators on the boundary, such
as the deformation states in equations \eqref{defo1}-\eqref{Tj}, the definition of the state
depends on the geometry used. That is to say, although such specifications are formally 
conformally invariant, the necessary regularisation procedure will break this symmetry.
When defined using operators requiring regularisation, conformally equivalent surfaces can
correspond to different states. This information is contained in the function which maps the
regulator from one coordinate system to another; here this was the function $s(\theta)$
which maps the regulator $\epsilon_z$ in the $z$-system to $\epsilon_\theta$ in the $\theta$-system.

\section*{Acknowledgements}
The author would like to thank K.~S.~Viswanathan and R.~C.~Rashkov
for enlightening discussions, and Simon Fraser University for a
Graduate Fellowship. This work was supported in part by the 
National Sciences and Engineering Research Council of Canada.

\setcounter{equation}{0}
\renewcommand{\theequation}{A-\arabic{equation}}
\section*{Appendix: Conformal Mapping to the Cone}
\begin{figure}[ht]
\includegraphics{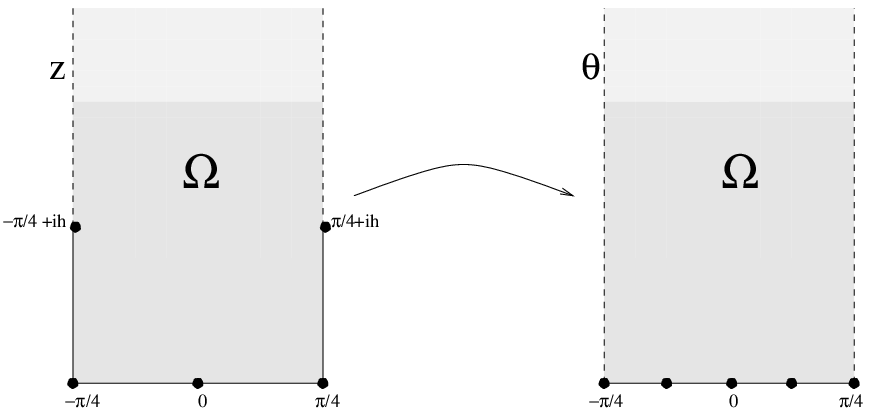}
\caption{\label{Map1} $\eta\cos 2\theta = \cos 2z$}
\end{figure}

Let $\Omega$ be the region of the complex plane given by $-\pi/4 < \re z < \pi/4$, $\im z>0$.
First let us construct a map from $\Omega$ to itself, which transforms the boundary 
$\partial\Omega$ as shown in \fgref{Map1}. 
That is, the contour given by line segments from
$z=-\pi/4+ih$ down to $z=-\pi/4$, across to $z=+\pi/4$ and up to $z=+\pi/4+ih$ should be mapped
to the segment of the real line $-\pi/4 < z < \pi/4$, with the rest of the boundary 
mapped accordingly. This map is given implicitly by the relation
\begin{equation}
\label{map1}
\cos 2\theta = \frac1\eta \cos 2z
,\end{equation}
where $\eta=\cosh 2h$.
Now, due to the periodicity of the functions in \eqref{map1}, this map in fact extends to
\begin{figure}[ht]
\includegraphics{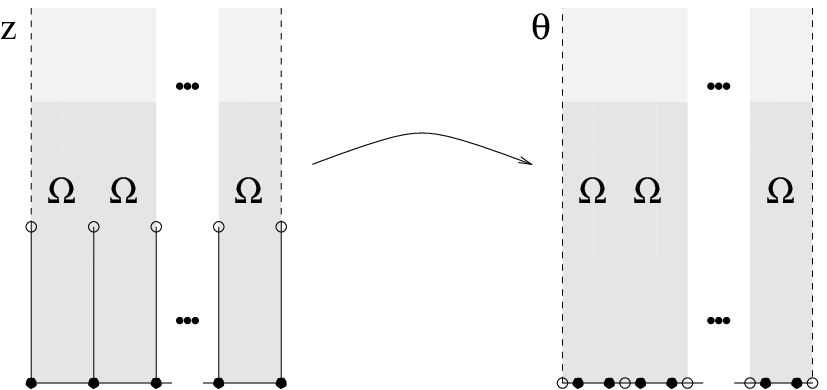}
\caption{\label{Map2} $\eta\cos 2\theta = \cos 2z$}
\end{figure}
arbitrarily many copies of the surfaces, as shown in \fgref{Map2}. Each `bucket' of width 
$\pi/2$ is folded down onto the real line. We will use the map for two copies of $\Omega$
for surfaces corresponding to $K_{11}$ and $K_{20}$, and the three-copy map for $K_{110}$ and $K_{200}$.
Explicitly, the map \eqref{map1} can of course be written as
\begin{equation}
\theta=\frac12\cos^{-1}\frac{\cos 2z}\eta
,\end{equation}
although we must be careful to note that the inverse cosine function is not single-valued.


\end{document}